\documentclass[aps,superscriptaddress, twocolumn,pre,nofootinbib]{revtex4-1}
\usepackage{amsmath,amssymb,amsfonts,amsthm}
\usepackage[english]{babel}
\usepackage{tabularx}
\usepackage{graphicx}
\usepackage{comment}
\usepackage{array}

\begin{document}

\title{Resource Efficient Gadgets for Compiling Adiabatic Quantum Optimization Problems}

\author{Ryan Babbush}
\affiliation{Department of Chemistry and Chemical Biology, Harvard University, Cambridge MA 02138 USA}

\author{Bryan O'Gorman}
\affiliation{Department of Chemistry and Chemical Biology, Harvard University, Cambridge MA 02138 USA}

\author{Al\'{a}n Aspuru-Guzik}
\email{Corresponding author Email: aspuru@chemistry.harvard.edu}
\affiliation{Department of Chemistry and Chemical Biology, Harvard University, Cambridge MA 02138 USA}

\date{\today }

\begin{abstract}
We develop a resource efficient method by which the ground-state of an arbitrary $k$-local, optimization Hamiltonian can be encoded as the ground-state of a $\left(k-1\right)$-local, optimization Hamiltonian. This result is important because adiabatic quantum algorithms are often most easily formulated using many-body interactions but experimentally available interactions are generally 2-body. In this context, the efficiency of a reduction gadget is measured by the number of ancilla qubits required as well as the amount of control precision needed to implement the resulting Hamiltonian. First, we optimize methods of applying these gadgets to obtain 2-local Hamiltonians using the least possible number of ancilla qubits. Next, we show a novel reduction gadget which minimizes control precision and a heuristic which uses this gadget to compile 3-local problems with a significant reduction in control precision. Finally, we present numerics which indicate a substantial decrease in the resources required to implement randomly generated, 3-body optimization Hamiltonians when compared to other methods in the literature.
\end{abstract}
\maketitle

\section{Introduction}

Our group has used quantum annealing to simulate classical problems of importance in chemistry such as lattice protein folding \cite{Perdomo-Ortiz2012, Perdomo2008,Babbush2012}. During the course of this work we developed tools, explained in this paper, which are essential for practically encoding and compiling classical problems into Hamiltonians suitable for experimental implementation. The adiabatic algorithm prepares a system in the ground-state of an arbitrary Hamiltonian through adiabatic evolution from the ground-state of a trivial Hamiltonian \cite{Farhi2001}. This strategy exploits the adiabatic theorem of quantum mechanics which states that a physical system remains in its instantaneous eigenstate if a given perturbation acts on it slowly enough and if there is a gap between the eigenvalue and the rest of the Hamiltonian's spectrum \cite{Born}. In 2004, this algorithm was shown to efficiently simulate any given quantum circuit and thus to be polynomially equivalent to standard quantum computation \cite{Aharonov2004}. That same year a proof by Kempe et al. demonstrated that adiabatic computation with a 2-local Hamiltonian accomplishes the same result \cite{Kempe2004}.

Though unlikely to be universal for adiabatic quantum computation, the 2-local quantum Ising model with 1-local transverse field has been realized using a wide array of technologies and is known to be a stoquastic Hamiltonian for which finding the ground-state is an \textsc{NP-Hard} problem \cite{Barahona1982, Bravyi2006,Jordan2010}. D-Wave System's current generation of quantum annealing machines are an important example of devices that implement this type of quantum Ising model Hamiltonian\footnote[2]{The D-Wave hardware implements a restricted form of the 2D Ising model, known as the Chimera graph, which can encode smaller instances of the general 2D Ising model. Compiling to this specific graph is beyond the scope of the present work.} \cite{Johnson2011a}. In the last two years, a large number of academic groups have used these annealing machines to solve a diversity of practical problems from protein folding to machine learning \cite{Neven2008a,Denchev2012,Bian2012,Hen2012,Boixo2013}. Unfortunately, classical optimization problems are most easily formulated using many-body interactions (many important problems, such as \textsc{3-XORSAT} have a natural 3-local encoding) \cite{Babbush2012,Farhi2012}. In order to reduce the locality of interactions from $k$-local to 2-local, one must employ the concept of a reduction gadget. A vigorous debate on the quantum nature of the D-Wave computer is currently underway \cite{Boixo2013,Wang2013,Smolin2013}. In this paper, we avoid that discussion and focus on the issue of efficiently constructing experimentally realizable Hamiltonians.

In their 2004 proof, Kempe et al. first introduced the notion of perturbative gadgets, which use perturbation theory to reproduce the low-energy subspace of an arbitrary, \textsc{QMA-Complete} $k$-local Hamiltonian with a $\left(k-1\right)$-local Hamiltonian in a larger Hilbert space, expanded by the addition of ancilla qubits \cite{Kempe2004,Jordan2008}. While this tool has been essential to a number of important proofs, none have used perturbative gadgets to efficiently encode practical problems into an experimentally realizable Hamiltonian \cite{Oliveira2005,Bravyi2006,Biamonte2007,Schuch2009}. One reason for this is that the perturbative approach requires the introduction of $k$ ancilla qubits for each term in a $k$-local Hamiltonian \cite{Jordan2008}. An even more significant problem with using perturbative gadgets for practical encodings is that each order of perturbation theory causes an exponential increase in the control precision required to implement the resulting Hamiltonian. In this context, control precision refers to the number of distinct field values that a device must be able to resolve in order to implement the requisite interactions in a given Hamiltonian. For instance, in an Ising model Hamiltonian with only 2-local, integer couplings, the control precision is defined as the magnitude of the largest coupler value divided by the greatest common factor of all couplings. D-Wave's newest device has 512 qubits and 4 bits of control precision, which amounts to 16 distinct values of coupler strength in both positive and negative biases. In practice, this means that problem size is often more limited by control precision than by qubits. Even for an ideal device which could implement couplings to arbitrary precision (but with finite maximum field strengths) we would want to avoid encodings which demand high control precision. This is because the separation of energy eigenstates is inversely proportional to the control precision, and thus the adiabatic runtime increases with control precision.

Fortunately, there exist a class of non-perturbative gadgets, formalized by Biamonte, which are significantly more efficient in terms of both ancilla and control precision  \cite{Biamonte2008}. We refer to these as \emph{exact classical gadgets} because they apply only for Hamiltonians in which all many-body terms are simultaneously diagonalizable, e.g. the D-Wave final Hamiltonian \cite{Biamonte2008}. This class of Hamiltonians can encode \textsc{NP-Hard} problems (but not \textsc{QMA Hard} unless \textsc{NP} = \textsc{QMA}) and is thus referred to as an optimization Hamiltonian \cite{Aharonov2004}. Exact gadgets work by substituting the product of two qubits in a $k$-local term with an ancilla qubit and introducing a 2-local penalty function which raises the energy of any state in which the product of the original two qubits is not equal to the state of the ancilla bit. This penalty function still raises control precision considerably but we show that the effect can be partially ameliorated by a different penalty function which uses additional ancilla qubits.

The central contribution of the present work is to introduce novel techniques for efficiently applying exact classical gadgets. In order to efficiently reduce locality, one must collapse many-body terms in a systematic fashion that takes into account the appearance of specific pairs of qubits in multiple higher-order terms. For applications in which qubits are the limiting resource, we demonstrate how to map the optimal reduction to \textsc{set cover} and \textsc{0-1 integer linear programming} (ILP) so that conventional solvers can be leveraged to quickly find the best encoding. For control precision limited problems we formalize the optimal reduction problem and propose a greedy algorithm that significantly outperforms the status quo. Finally, we present numerics which demonstrate the significant advantage of using these optimized gadgets and gadget application techniques over all previously mentioned reduction techniques in the literature.

\section{Optimal reduction gadgets}
In order to compile \textsc{NP-hard} optimization problems into an experimental Hamiltonian, one must encode the problem of interest into a graph of binary variables with physically realizable interactions. Perhaps the simplest model of interacting binary variables is \textsc{Polynomial Unconstrained Binary Optimization} (\textsc{PUBO}): given a pseudo-Boolean function $f:\mathbb B^N \rightarrow \mathbb R$, find an assignment $\mathbf{x} \in \mathbb B^N$ such that $f\left(\mathbf{x}\right) = \text{min}\left[f\left(\mathbb B ^N\right)\right]$, where $\mathbb B = \left\{ 0, 1 \right\}$. Every pseudo-Boolean $f$ has a unique multi-linear polynomial representation
\begin{equation}
f\left(\mathbf{x}\right) = \sum_{S \subseteq \lbrace 1, \cdots, N \}} c_S \prod_{i \in S} x_i,
\end{equation}
where $c_S\in \mathbb R$. From this expression we can construct an optimization Hamiltonian that embeds the energy landscape of a given PUBO in its eigenspectrum,
\begin{equation}
H\left(f\right) = \sum_{S \subseteq \{ 1, \cdots, N \}} c_S \prod_{i \in S} q_i,
\end{equation}
acting on $N$ qubits, where $q_i=  \frac{1}{2} \left(I^{\otimes N}-Z_i\right)$ and $Z_i$ is the Pauli matrix $\sigma^z$ acting on the $i$th qubit, i.e.
\begin{equation}
Z_i= I^{\otimes (i-1)}\otimes \sigma^z \otimes I^{\otimes (N-i)},
\end{equation}
where $I$ is the one-qubit identity operator. Note that while we write  $H\left(f\right)$ for convenience, in practice $f$ will be specified by its coefficients $c_S$. Every element $|\mathbf x\rangle$ of the computational basis is an eigenstate of $H\left(f\right)$ with eigenvalue $f\left(\mathbf x\right)$. Specifically, the ground state of $H\left(f\right)$ is spanned by the set of states $|\mathbf x\rangle$ such that $f\left(\mathbf x\right) = \text{min}\left[f\left(\mathbb B^N\right)\right]$.

However, experimental interactions are typically limited to pairwise couplings between qubits, allowing Hamiltonians of the form\footnote[3]{In our definition, the case in which the indices are equal is used to include 1-local terms: $q_iq_i=q_i$.}
\begin{equation}
\label{eq:2-localHamiltonian}
H\left(f\right) =  \sum_{1 \leq i \leq j \leq N} \alpha_{ij} q_i q_j,
\end{equation} 
where $\alpha_{ij} \in \mathbb R$. Such Hamiltonians correspond to a second-order pseudo-Boolean $f$,
\begin{equation}
\label{eq:QUBO}
f\left(\mathbf{x}\right) =  \sum_{1 \leq i \leq j \leq N} \alpha_{ij} x_i x_j.
\end{equation} 
Thus, to encode a general instance of \textsc{PUBO} into an experimentally realizable Hamiltonian, one must reduce the problem to \textsc{Quadratic Unconstrained Binary Optimization} (\textsc{QUBO}), defined analogously to \textsc{PUBO} with the restriction that the pseudo-Boolean function to be minimized is quadratic. In practice, many common optimization problems have been reduced to \textsc{PUBO} in such a way that the pseudo-Boolean function to be minimized is cubic, i.e. of the form
\begin{equation}
f\left(\mathbf x\right) = \sum_{1\leq i \leq j \leq N} \alpha_{ij} x_i x_j + \sum_{1\leq i < j < k\leq N} \alpha_{ijk} x_i x_j x_k.\nonumber
\end{equation}
It is therefore desirable to have a general method for reducing a cubic function $f:\mathbb B^N \rightarrow \mathbb R$ to a quadratic function $f':\mathbb B^{N'} \rightarrow \mathbb R$ in such a way that an assignment $\mathbf{x}\in \mathbb B^N$ that minimizes $f$ can be efficiently computed given an assignment $\mathbf{x'}$ that minimizes $f'$, where $N'$ is a polynomial function of $N$. One family of methods employs a set of $N'-N$ ancilla variables $\{ y_1, \cdots, y_{N'-N} \} \in \mathbb B^{N'-N}$ such that if $\left(x_1, \cdots, x_N, y_1, \cdots y_{N'-N}\right)$ minimizes $f'$, then $\left(x_1, \cdots, x_N\right)$ minimizes $f$. That is, a minimizing assignment $\left(x_1, \cdots, x_N\right)$ of $f$ is directly encoded in the $N$ computational qubits of a ground state $|x_1 \cdots x_N y_1 \cdots y_{N'-N}\rangle$ of $H\left(f'\right)$. In the methods examined here, each ancilla variable corresponds to a pair of computational variables $\left(i,j\right)$ and so for convenience is denoted by $x_{ij}$ or $x_{ij}^{(m)}$.

\subsection{Minimal ancilla gadget} 
Integral to the exact gadget is the penalty function
\begin{equation}
s\left(x, y, z\right) = 3 z + x y - 2xz - 2yz,
\end{equation}
with the important property that $s\left(x,y,z\right)=0$ if $xy=z$ and $s\left(x,y,z\right) \geq 1$ if $xy \neq z$, as shown in Table 1 \cite{Biamonte2008}. While $s$ is not the only quadratic ternary pseudo-Boolean with this property, we will show that it is optimal for our purposes.
\begin{table}[h]
\caption{Truth table for ancilla gadget}
\begin{tabularx}{\columnwidth}{>{\centering}m{0.225\columnwidth} >{\centering}m{0.225\columnwidth}>{\centering}m{0.225\columnwidth}>{\centering}m{0.225\columnwidth}}
\hline\hline
\noalign{\vskip 1mm}
x & y & z & $s\left(x,y,z\right)$ \tabularnewline
\noalign{\vskip 1mm} 
\hline
\noalign{\vskip 2mm}    
0 & 0 & 0 & 0\tabularnewline
0 & 1 & 0 & 0\tabularnewline
1 & 0 & 0 & 0\tabularnewline
1 & 1 & 1 & 0\tabularnewline
0 & 0 & 1 & 3\tabularnewline
0 & 1 & 1 & 1\tabularnewline
1 & 0 & 1 & 1\tabularnewline
1 & 1 & 0 & 1\tabularnewline
\noalign{\vskip 1mm}    
\hline\hline
\end{tabularx}
\end{table}

In our reductions, we replace a part $x_ix_j$ of a 3-local term $x_ix_jx_k$ with $x_{ij}$, where $x_{ij}$ is an ancilla variable, thereby reducing locality, while simultaneously adding the penalty function $s\left(x_i,x_j,x_{ij}\right)$, scaled by an appropriate factor to ensure that the value of the reduced form is greater if $x_{ij}\neq x_i x_j$ than it is if $x_{ij}=x_ix_j$, for any assignment of the computational variables. In this way, we ensure that if an assignment of the computational and ancilla variables minimizes the reduced form, then that assignment of the computational variables also minimizes the original form. Consider the reduction
\begin{equation}
\label{eq:1gadget}
\alpha_{ijk}x_ix_jx_k \rightarrow \alpha_{ijk} x_{ij} x_k + \left(1+|\alpha_{ijk}|\right)s\left(x_i, x_j, x_{ij}\right).
\end{equation}
If $x_{ij}=x_i x_j$, then $s\left(x_i, x_j, x_{ij}\right) = 0$ and the reduced form simplifies to the unreduced form $\alpha_{ijk}x_i x_j x_k$. If $x_{ij}=1 - x_i x_j$, then $s(x_i, x_j, 1 - x_ix_j) = 3 - 2x_i -2x_j +2x_ix_j$ and the reduced form always has a greater value than it does if $x_{ij}=x_ix_j$. That is,
\begin{align}
&\alpha_{ijk}\left(1-x_ix_j\right)x_k+\left(1+|\alpha_{ijk}|\right)s\left(x_i,x_j,1-x_{i}x_j\right) \\
&>\alpha_{ijk}\left(x_ix_j\right)x_k + \left(1 + |\alpha_{ijk}|\right) s\left(x_i, x_j, x_i x_j\right) \nonumber\\ 
&= \alpha_{ijk}x_ix_jx_k \nonumber
\end{align}
for all $x_i$, $x_j$, and $x_k$.
To decrease the number of ancilla variables needed to reduce many 3-local terms, it is advantageous to use the same ancilla variable $x_{ij}$ to reduce more than one 3-local term. Let $K_{ij}$ be the set of indices $k$ such that the term $x_i x_j x_k$ is reduced using the ancilla variable $x_{ij}$ corresponding to the pair of variables $\left\{x_i, x_j\right\}$. Each non-zero 3-local term is reduced using exactly one ancilla, and so we must choose $\{ K_{ij}\}$ such that for each $\alpha_{ijk}\neq 0$, there is exactly one pair of indices $\{ w, v\rbrace$ with $\{ w, v, K_{wv} \rbrace = \{ i, j, k\rbrace$\footnote[4]{Note that the indices on the coefficients are unordered, e.g. $\alpha_{ijk}=\alpha_{kji}=\alpha_{jki}$.}. Then the entire set of 3-local terms can be reduced by
\begin{align}
\label{eq:simpleMultipleReduction}
&\sum_{1\leq i < j < k \leq N}\alpha_{ijk}x_ix_jx_k = \sum_{1\leq i < j \leq N}\sum_{k\in K_{ij}}\alpha_{ijk}x_ix_jx_k\\
&\rightarrow \sum_{1\leq i \leq j \leq N}\sum_{k \in K_{ij}} \left(\alpha_{ijk}x_{ij}x_k  
+\left(1+|\alpha_{ijk}|\right)s\left(x_i, x_j, x_{ij}\right)\right) \nonumber,
\end{align} 
where the single term reduction in Eq.~\eqref{eq:1gadget} is applied to every term in the rewritten original expression. The essential conditions (that, for any $i$ and $j$ for which an ancilla variable is used, the value of the reduced form is greater if $x_{ij}\neq x_ix_j$ than the value thereof if $x_{ij}=x_ix_j$ and in the latter case the reduced form is equal to the original form) are preserved by linearity. In Section 3, we explain a method for choosing which pair of variables to use to reduce each 3-local term (i.e. for choosing $K_{ij}$ with the constraints given) in a way that minimizes the total number of ancilla variables (the number of non-empty $K_{ij}$). In the Appendix we generalize this strategy to minimize the number of ancilla required in 4-local to 2-local reductions.

\subsection{Minimal control precision}
It is often the case that the limiting factor in encoding a \textsc{PUBO} instance into experimentally realizable form is the control precision rather than the number of qubits available \cite{Babbush2012}. Existing hardware is able to implement 2-local Hamiltonians of the form in Eq.~\eqref{eq:2-localHamiltonian} such that the coefficients are integral multiples of a fixed step size $\Delta_{\alpha}$ with a maximum magnitude of $N_{\alpha}\Delta_{\alpha}$, where $N_{\alpha}$ is the control precision. An arbitrary 2-local Hamiltonian can be made to have coefficients that are integral multiples of $\Delta_{\alpha}$ by dividing them all by their greatest common divisor and multiplying by $\Delta_{\alpha}$. The control precision needed for an arbitrary instance is thus the quotient of the greatest magnitude of the coefficients and their greatest common divisor. We assume without loss of generality that the coefficients of the \textsc{PUBO} to be reduced are integers and structure the reductions so that the reduced \textsc{QUBO} also has integral coefficients. The greatest common divisor of the coefficients of the reduced \textsc{QUBO} is thus one with high probability, and the control precision needed is the greatest magnitude of the coefficients. As a preliminary, we show that $s$ as defined is optimal in that the greatest coefficient (3) cannot be reduced any further.

Suppose $f\left(x_1, x_2, x_3\right)$ is a quadratic pseudo-Boolean function with integer coefficients (i.e. in the form of Eq.~\eqref{eq:QUBO}) such that $f\left(x_1, x_2, x_3\right)=0$ if $x_3=x_1x_2$ and is at least one otherwise. First note that $f\left(0,0,0\right)=0$ and thus that $f\left(1,0,0\right)= \alpha_{11}=0$ and $f\left(0,1,0\right)=\alpha_{22}=0$. Because $f\left(1,1,1\right)=\alpha_{33}+\alpha_{12}+\alpha_{13}+\alpha_{23}=0$, $\alpha_{33}+\alpha_{23}=-\alpha_{12}-\alpha_{13}$, and so $f\left(0,1,1\right)=\alpha_{33}+\alpha_{23}=-\alpha_{12}-\alpha_{13}\geq 1$, which implies $\alpha_{13}\leq -\alpha_{12}-1$. Because $\alpha_{12}=f\left(1,1,0\right)\geq 1$, $\alpha_{13}\leq -2$. Finally, $f\left(1,0,1\right)=\alpha_{33}+\alpha_{13}\geq 1$ and so $\alpha_{33}\geq 1-\alpha_{13}\geq 3$.

For each $i$ and $j$ in the reduction shown in Eq.~\eqref{eq:simpleMultipleReduction}, $\left(1+|\alpha_{ijk}|\right)s\left(x_i, x_j, x_{ij}\right)$ is added for each $k\in K_{ij}$, and so the coefficients in $s\left(x_i, x_j, x_{ij}\right)$ are multiplied by $\sum_{k\in K_{ij}}\left(1+|\alpha_{ijk}|\right)$. We claim that this factor can be decreased using the reduction
\begin{align}
\label{eq:group}
& \sum_{k\in K_{ij}}\alpha_{ijk}x_ix_jx_k \rightarrow \sum_{k\in K_{ij}}\alpha_{ijk}x_{ij}x_k + \delta_{ij}s\left(x_i, x_j, x_{ij}\right),\\
&\text{where}\nonumber\\
& \delta_{ij}=1 + \text{max}\left\{ \sum_{k \in \{ k \in K_{ij} | \alpha_{ijk} > 0\}}\alpha_{ijk},
\sum_{k \in \{ k \in K_{ij} | \alpha_{ijk} < 0\}}\!\!\!- \alpha_{ijk}\right\},\nonumber
\end{align}
for each $i$ and $j$. For all $x_i$, $x_j$, and $\{ x_k | k \in K_{ij}\}$,
\begin{align}
& \sum_{k\in K_{ij}}\alpha_{ijk}(1-x_i x_j)x_k + \delta_{ij}s\left(x_i, x_j, 1 - x_i x_j \right) \\
& > \sum_{k\in K_{ij}}\alpha_{ijk}(x_i x_j)x_k + \delta_{ij}s\left(x_i, x_j, x_i x_j\right). \nonumber
\end{align}
That is, for any assignment of the computational variables, the value of the reduced form is greater if the ancilla variable $x_{ij}\neq x_ix_j$ than it is if $x_{ij}=x_i x_j$.  The $\delta_{ij}$ given in Eq.~\eqref{eq:group} is optimal in the sense that it requires the least control precision of all possibilities which satisfy the appropriate conditions. Consider the reduced form
\begin{equation}
x_{ij}\sum_{k\in K_{ij}}\alpha_{ijk}x_k+\delta s(x_i,x_j, x_{ij})
\end{equation}
for some $\delta \in \mathbb{Z}$ to be determined. We must guarantee that
\begin{align}
&(1-x_ix_j)\sum_{k\in K_{ij}}\alpha_{ijk}x_k+\delta s(x_i,x_j,1-x_ix_j) \\
& > x_ix_j\sum_{k\in K_{ij}}\alpha_{ijk}x_k+\delta s(x_i,x_j, x_ix_j) \nonumber
\end{align}
for all $x_i$, $x_j$, and $\{ x_k|k\in K_{ij}\}$. For $x_i=1$ and $x_j=0$ or $x_i=0$ and $x_j=1$, this inequality simplifies to
\begin{equation}
\delta > - \sum_{k \in K_{ij}}\alpha_{ijk}x_k,
\label{eq:xi!=xj}
\end{equation}
for $x_i=x_j=1$ it simplifies to
\begin{equation}
\delta > \sum_{k \in K_{ij}}\alpha_{ijk}x_k,
\label{eq:xi=xj=1}
\end{equation}
and for $x_i=x_j=0$ it simplifies to
\begin{equation}
\delta > -\frac{1}{3}\sum_{k\in K_{ij}}\alpha_{ijk}x_k.
\label{eq:xi=xj=0}
\end{equation}
Eq.~\eqref{eq:xi=xj=0} is implied by Eq.~\eqref{eq:xi!=xj} and so it is sufficient to ensure that Eq.~\eqref{eq:xi!=xj} and Eq.~\eqref{eq:xi=xj=1} are satisfied. We see that the term $-\sum_{k \in K_{ij}} \alpha_{ijk} x_k$ is greatest when 
\begin{equation}
\label{cond}
x_k = \begin{cases} 
1 & \text{if } \alpha_{ijk} < 0\\
0 & \text{if } \alpha_{ijk} > 0
\end{cases},
\end{equation}
 and so if and only if
\begin{equation}
\delta > \sum_{k \in \{ k \in K_{ij}| \alpha_{ijk} < 0\}} \!\!\!-\alpha_{ijk},
\label{eq:sumNeg}
\end{equation}
 then Eq.~\eqref{eq:xi!=xj} is satisfied for all $\{ x_k | k \in K_{ij} \}$. The term $\sum_{k \in K_{ij}} \alpha_{ijk} x_k$ is greatest under the exact opposite conditions as Eq.~\eqref{cond}. Thus, if and only if
\begin{equation}
\delta > \sum_{k \in \{ k \in K_{ij} | \alpha_{ijk} > 0\}}\alpha_{ijk}
\label{eq:sumPos}
\end{equation}
then Eq.~\eqref{eq:xi=xj=1} is satisfied for all $\{ x_k | k \in K_{ij} \}$. Together, Eq.~\eqref{eq:sumNeg} and Eq.~\eqref{eq:sumPos} and imply that
\begin{equation}
\delta >\text{max}\left\{ \sum_{k \in \{ k \in K_{ij} | \alpha_{ijk} > 0\}}\alpha_{ijk},
\sum_{k \in \{ k \in K_{ij} | \alpha_{ijk} < 0\}}\!\!\!- \alpha_{ijk}\right\}.
\end{equation}
Note that the terms introduced in Eq.~\eqref{eq:group} only appear in the reduction for that pair $(i,j)$, and so the coefficient for a term therein is the coefficient in the total reduced form, with the exception of $x_ix_j$ which may also appear in the original unreduced form, which is to be addressed later. The greatest term introduced in Eq.~\eqref{eq:group} is $3\delta_{ij}$, which greatly increases the control precision needed.

Below, we introduce an alternative method that adds terms whose greatest coefficient is approximately a third of this. Because the complexity of the final form obscures the simplicity of the method, we begin with a special case and extend it gradually to the general case. To reduce a single term whose coefficient is divisible by three, we introduce three ancillary bits and penalty functions:
\begin{align}
\alpha_{ijk}x_i x_j x_k & \rightarrow \frac{\alpha_{ijk}}{3}\left(x_{ij}^{(1)}+x_{ij}^{(2)}+ x_{ij}^{(3)}\right)x_k \\
& +\left(1+\left|\frac{\alpha_{ijk}}{3}\right|\right)\sum_{m=1}^3 s\left(x_i,x_j,x_{ij}^{(m)}\right)\nonumber
\end{align}
When $x_{ij}^{(1)}=x_{ij}^{(2)}=x_{ij}^{(3)}=x_ix_j$, the reduced form simplifies to $\alpha_{ijk}x_ix_jx_k$. Otherwise, it is always greater than $\alpha_{ijk} x_i x_j x_k$, and so the reduction is valid. Furthermore, the greatest coefficient introduced is $3+|\alpha_{ijk}|$. In general however, the coefficient will not be divisible by 3. In that case, we define a new coefficient $\beta_{ijk}^{(m)}$ for each ancilla variable $x_{ij}^{(m)}$ that depends on $\alpha_{ijk} \text{ mod } 3$ such that each $\beta_{ijk}^{(m)}$ is an integer and $\sum_{m=1}\beta_{ijk}^{(m)}=\alpha_{ijk}$. This is elucidated by Table 2.
\begin{table}[h]
\caption{Integer coefficients so that $\sum_{m=1}\beta_{ijk}^{(m)}=\alpha_{ijk}$}
\begin{tabularx}{\columnwidth}{>{\centering}m{0.225\columnwidth} >{\centering}m{0.225\columnwidth}>{\centering}m{0.225\columnwidth}>{\centering}m{0.225\columnwidth}}
\hline\hline
\noalign{\vskip 1mm}
$\alpha_{ijk} \text{ mod } 3$ & $ \beta_{ijk}^{(1)} $ &  $\beta_{ijk}^{(2)} $ &  $\beta_{ijk}^{(3)}$ \tabularnewline
\noalign{\vskip 1mm} 
\hline
\noalign{\vskip 2mm}    
0 & $\alpha_{ijk}/3$ & $\alpha_{ijk}/3$ & $\alpha_{ijk}/3$\tabularnewline
1 & $\left(\alpha_{ijk}+2\right)/3$ &  $\left(\alpha_{ijk}-1\right)/3$ &  $\left(\alpha_{ijk}-1\right)/3$\tabularnewline
2 &  $\left(\alpha_{ijk}+1\right)/3$ &  $\left(\alpha_{ijk}+1\right)/3$ &  $\left(\alpha_{ijk}-2\right)/3$\tabularnewline
\noalign{\vskip 1mm}    
\hline\hline
\end{tabularx}
\end{table}
We now use the reduction
\begin{align}
\alpha_{ijk}x_ix_jx_k & \rightarrow \left(\beta_{ijk}^{(1)}x_{ij}^{(1)}+\beta_{ijk}^{(2)}x_{ij}^{(2)} +\beta_{ijk}^{(3)}x_{ij}^{(3)}\right)x_k\\
& + \sum_{m=1}^3\left(1+\left|\beta_{ijk}^{(m)}\right|\right)s\left(x_i,x_j,x_{ij}^{(m)}\right) \nonumber.
\end{align}
If $x_{ij}^{(1)}=x_{ij}^{(2)}=x_{ij}^{(3)}=x_i x_j$, then $s(x_i,x_j,x_{ij}^{(m)})=0$ and this simplifies to $\alpha_{ijk}x_i x_j x_k$.  
We can rewrite the replacement terms as,
\begin{equation}
\sum_{m=1}^{3}\left(\beta_{ijk}^{(m)}x_{ij}^{(m)}x_k+\left(1+\left|\beta_{ijk}\right|\right)s\left(x_i,x_j,x_{ij}\right)\right)
\end{equation}
In all cases and for each $m$
\small
\begin{equation}
\beta_{ijk}^{(m)}x_{ijk}^{(m)}x_k+\left(1+\left|\beta_{ijk}\right|\right)s(x_i,x_j,x_{ij}^{(m)}) \geq \beta_{ijk}^{(m)}x_i x_j x_k.
\end{equation}
If not $x_{ij}^{(1)}=x_{ij}^{(2)}=x_{ij}^{(3)}=x_i x_j$, strict inequality holds for at least one $m$ and the replacement terms are greater than $\alpha_{ijk}x_i x_j x_k$.
Here, the greatest coefficient is 
\begin{equation}
3+\max\left\{ 3\left|\beta_{ijk}^{(m)}\right|,\left|\alpha_{ijk}\right|\right\}.
\end{equation}
Finally, we use the same set of ancilla variables $\left\{\alpha_{ij}^{(m)}\right\}$ to reduce all of the 3-local terms:
\begin{align}
& \sum_{1\leq i < j < k \leq N}\alpha_{ijk}x_ix_jx_k 
= \sum_{1\leq i < j \leq N} \sum_{k\in K_{ij}}
  \sum_{m=1}^3 \beta_{ijk}^{(m)} x_i x_j x_k \\
& \rightarrow \sum_{1\leq i \leq j \leq N} \sum_{m=1}^3 \left(\sum_{k\in K_{ij}} \beta_{ijk}^{(m)} x_{ij}^{(m)} x_k + \delta_{ij}^{(m)}s(x_i, x_j, x_{ij}^{(m)})\right),\nonumber\\
&\text{where}\nonumber\\
& \delta_{ij}^{(m)} = 1 + \text{max}\left\{\sum_{k \in \{ k \in K_{ij} | \beta_{ijk}^{(m)} > 0 \}} \beta_{ijk}^{(m)},
\sum_{k \in \{ k \in K_{ij} | \beta_{ijk}^{(m)} < 0 \}} \!\!\! -\beta_{ijk}^{(m)}\right\}\nonumber
\end{align}
and $K_{ij}$ is defined as above with the same constraints. In the reduced form, for every $i$, $j$, and $m$ the coefficient of $x_{ij}^{(m)}$ is  $3\delta_{ij}^{(m)}$ and for every $i$ and $j$ the coefficient of $x_ix_j$ is $\sum_{m=1}^3\delta_{ij}^{(m)}$. The latter will be added to the coefficient $\alpha_{ij}$ of the corresponding quadratic term in the original expression. Thus the control precision needed is
\begin{equation}
\operatorname*{min}_{\{ K_{ij}\}} \left(\text{max}\left\{
\operatorname*{max}_{i,j,m} \left(3\delta_{ij}^{(m)} \right), 
\operatorname*{max}_{i,j} \left|\alpha_{ij}+\sum_{m=1}^3 \delta_{ij}^{(m)}\right|
\right\}\right).
\label{eq:controlPrecision}
\end{equation}
In Section 3 we describe a greedy algorithm to find a set of $K_{ij}$ that greatly decreases the control precision needed.

\section{Efficient encoding techniques}
With the exception of the 3-ancilla gadget to reduce control precision, the classical gadgets we have described have already been characterized in the literature. However, knowing these formulas is not enough to efficiently encode a problem. In the following two sections we describe how to efficiently apply these gadgets so that the resulting Hamiltonian meets the demands of available hardware. For simplicity, and because it is the most frequently encountered situation, we will focus on reductions from 3-local to 2-local. We also describe the 4-local to 2-local reduction in the Appendix.

When working with a qubit limited encoding, the goal in applying these gadgets will be to choose the smallest set of qubit pairs that collapses all 3-local terms. We explain how to cast this problem as canonical \textsc{set cover} and map to \textsc{0-1 ILP} so that popular optimization software can be leveraged to find the  \emph{optimal} set of collapsing pairs. When working with a control precision limited encoding, the goal is to choose the set of qubits for which the sum of penalty functions contains the smallest maximum coefficient. We approach this problem with a greedy algorithm but later show numerics which validate the efficiency of our technique.

\subsection{Limited ancilla reduction technique}

The qubit-optimized application of classical gadgets can be cast as \textsc{set cover}. In this context, the universe $U$ that we seek to cover is the set of 3-local terms that we must collapse. For example, $U = \left\{ x_1 x_2 x_3, x_1 x_4 x_5, x_2 x_3 x_5\right\}$. Treating each 3-local term as a set of single qubits, we define $A$ as the union of all 2-subsets of each 3-local term. In the example given,
\begin{align}
A & =\bigcup _{i=1}^{\left\vert{U}\right\vert} \left\{X \mid X \in 2^{U_i} \wedge {\left\vert{X}\right\vert} = 2 \right\}\\
& = \left\{x_1x_2, x_1 x_3, x_2 x_3\right\} \cup \left\{x_1 x_4,x_1 x_5, x_4 x_5\right\} \cup \left\{x_2 x_3, x_2 x_5, x_3 x_5\right\}\nonumber\\
& =  \left\{x_1x_2, x_1 x_3, x_1 x_4, x_1 x_5, x_2 x_3, x_2 x_5, x_3 x_5, x_4 x_5 \right\}.\nonumber
\end{align}
Next, we construct $S$ by replacing each element $A_i$ with the union of proper supersets of $A_i$ in $U$,
\begin{align}
S & = \bigcup _{i=1}^{\left\vert{A}\right\vert} \left\{\left\{ X \mid X \in U \wedge X \supsetneq A_i \right\}\right\}\\
& = \left\{ \left\{x_1 x_2 x_3\right\}, \left\{x_1 x_2 x_3\right\}, \left\{x_1 x_4 x_5\right\}, \left\{x_1 x_4 x_5\right\}, \right. \nonumber\\
& \qquad \left.\left\{x_1 x_2 x_3,x_2 x_3 x_5\right\}, \left\{x_2 x_3 x_5\right\}, \left\{x_2 x_3 x_5 \right\}, \left\{x_2 x_3 x_5\right\}\right\}.\nonumber
\end{align}

In this way, $A$ is the set of products of pairs of qubits $x_ix_j$ that can be used in the reduction, and each element $S_i$ is the set of 3-local terms that the corresponding $A_i$ can be used to reduce. The problem is clearly $\textsc{set cover}$ if we view the 3-local terms as elements (as opposed to sets themselves). Given $U$ and $S$, find the minimal covering set, i.e. $ \displaystyle \operatorname*{argmin}_{\left\{ C \mid C \subseteq S \wedge \bigcup C = U \right\}} |C| $. In this form, the problem is easily cast as \textsc{0-1 ILP}. \textsc{0-1 ILP} is the problem of finding a Boolean-valued vector $v$ that minimizes the quantity $c^{\textrm{T}} v$ subject to $M v \geq b$. In \textsc{set cover} each element of $v$ is a Boolean which says whether or not to include the associated element of $S$ in the cover $C$. Thus, $c$ is a vector of ones with length equal to the cardinality of $S$ so that the cost function $c^{\textrm{T}}v$ represents the cardinality of $C$.

\begin{figure}[h!]
\includegraphics[width=\columnwidth]{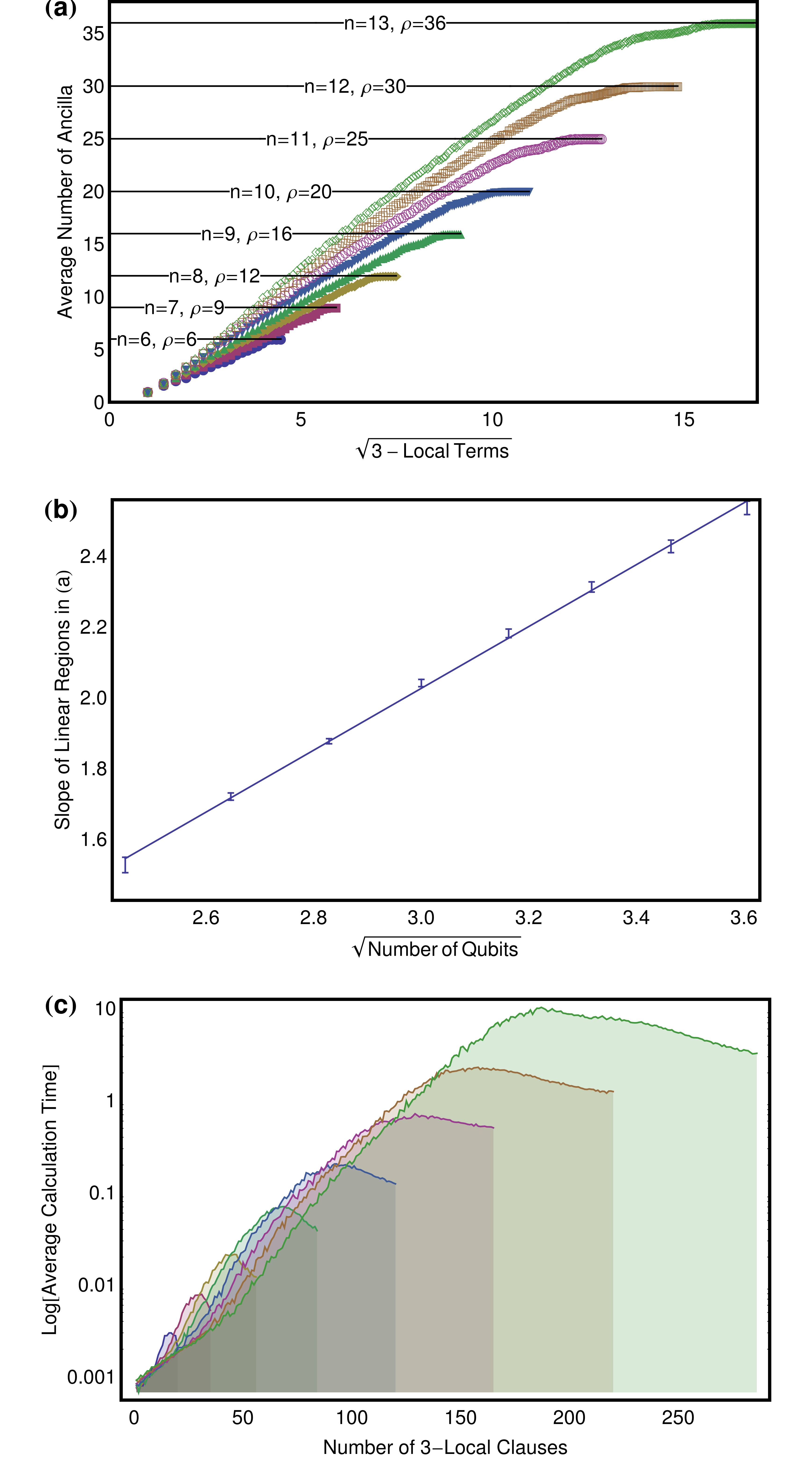}
 \caption{\label{ancilla} Performance of ancilla reduction scheme computed with \emph{Mathematica 9} ILP solver. Numerics were collected on randomly generated 3-local polynomial unconstrained binary optimization (\textsc{PUBO}) with $n$ logical qubits and $\lambda$ 3-local clauses. For each value of $n$, data were collected from 1,000 instances at every possible value of $\lambda$, i.e. $\left\{ \lambda \in \mathbb{Z} \mid 1 \leq \lambda \leq C^n_3\right\}$. Different colors indicate different values of $n$ ranging between 6 and 12 qubits. 1a: average number of ancilla required for reduction to 2-local versus $\sqrt{\lambda}$. 1b: slope of fits to linear region of aforementioned plot as a function of $\sqrt{n}$. Linear fits in top plots indicate that ancilla requirements scale as $\sqrt{n\lambda}$ until saturating at $\rho\left(n\right)$. 1c: semi-log plot showing average time in seconds for ILP to solve random instance.}
\end{figure}

The matrix $M$ multiplies $v$ to set up a system of equations which guarantees that $C$ covers $U$. Thus, the matrix element $M_{ij}$ is 1 if the $S_j$ contains the $U_i$ and 0 otherwise. Accordingly $b$ is a vector of all ones with length equal to the cardinality of $U$. Both $\textsc{set cover}$ and $\textsc{0-1 ILP}$ are well known to be \textsc{NP-Complete}. In fact, the exact problem of cubic to quadratic polynomial binary reduction has been shown to be \textsc{NP-Complete} by analogy with vertex cover \cite{Boros2002}.

In Figure 1 we show numerics that demonstrate the efficiency of embeddings that make use of this reduction technique. For the case of 3-local to 2-local PUBO reduction, the complexity of a random problem instance is characterized by the number of logical qubits, $n$, and the number of 3-local clauses, $\lambda$. While ancilla requirements scale as $3 \lambda$ for perturbative gadgets and $\lambda$ for exact gadgets without optimized application, numerics from Figure 1a and 1b indicate that our ancilla requirements scale as $\sqrt{n \lambda}$ until reaching an asymptote equal to the quarter squares function, defined as $ \rho\left(n\right) = \left\lfloor \frac{\left(n-1\right)^2}{4} \right\rfloor $. A proof of this bound is shown in the Appendix. In terms of the clause to variable ratio, $r = \lambda / n$, we see that our method scales as $n \sqrt{r}$ whereas the other methods scale as $3 n r$ and $n r$, respectively. Thus, we see a quadratic improvement in the number of ancilla required for a given clause to variable ratio but after a certain point, our method saturates and requires no additional ancilla, representing an undefined improvement over other methods.

Unfortunately, we should not expect to do better than a quadratic improvement for extremely large problem sizes because the constant scaling region appears to coincide with the most difficult to reduce problem instances as indicated by the computational time scaling in Figure 1c. In this regime, exact ILP solvers might take exponentially long to find the minimal cover. The worst case scenario is that the integrality gap of the ILP scales with the logarithm of $\lambda$, which would preclude the existence of a polynomial-time relaxation algorithm to approximate the solution beyond a logarithmic factor \cite{Arora1998}. There does not seem to be any clear connection between the complexity of a \textsc{PUBO} instance and the complexity of reducing that instance to \textsc{QUBO}; thus, we should have no reason to suspect that an average, hard \textsc{PUBO} problem will take exponential time to reduce to \textsc{QUBO} with ILP. However, for intractably large instances in the difficult clause to variable ratio regime, there exist greedy algorithms in the literature, for instance \textsc{ReduceMin} in Section 4.4 of \cite{Boros2002} which finds the pair of indices that appears most in qubit triplets, reduces that pair in all 3-local terms, and repeats this process until all triplets are depleted.

\subsection{Limited control precision reduction technique}

To minimize the control precision, as expressed in Eq.~\eqref{eq:controlPrecision}, we develop a greedy algorithm which chooses the collapsing pairs, $\{ K_{ij}\}$. 
Recall that $K_{ij}$ is the set of indices $k$ such that the term $x_i x_j x_k$ is reduced using the ancilla variable $x_{ij}$ corresponding to the pair of variables $\left(x_i, x_j\right)$. In the following pseudo-code we employ the convention that $K(\{ i, j\})=K_{ij}$, $\alpha(\{ i, j, k\rbrace)= \alpha_{ijk}$, and $\alpha(\{ i, j\rbrace )=\alpha_{ij}$ for ease of exposition.\\

\begin{figure}[h]
\framebox[\linewidth][l]{
\parbox{\linewidth}{
\raggedright
\parindent = 10pt
\textsc{Input}: $N$, $\alpha: \{\{i,j,k\})|1\leq i< j < k\leq N\} \rightarrow \mathbb R$ \\
\textsc{Initialization}: \\
\indent \textbf{for} $1\leq i < j \leq N$:\\
\indent \indent $K(\{ i, j\}) =  \emptyset$\\
\indent $A = \left\{ \{ i, j, k \} | 1 \leq i < j < k \leq N \wedge \alpha(\{ i, j, k\rbrace) \neq 0 \right\rbrace$\\
\indent \textbf{for} $a \in A$:\\
\indent \indent $B(a) = \left\{ \{ p, q \} | \{ p, q \rbrace \subset a \right\rbrace$\\
\textsc{Loop}:\\
\indent \textbf{while} $|A| > 0$:\\
\indent \indent \textbf{for} $a \in A$:\\
\indent \indent \indent \textbf{for} $b \in B(a)$:\\
\indent \indent \indent \indent $\Theta = \left\{ \alpha(b\cup \{ k\}) | k \in K(b)\right\rbrace \cup \{ \alpha(a)\rbrace$\\
\indent \indent \indent \indent $\Theta^+ = \left\{ \theta | \theta \in \Theta \wedge \theta > 0\right\}$\\ 
\indent \indent \indent \indent $\Theta^- = \left\{ \theta | \theta \in \Theta \wedge \theta < 0\right\}$\\ 
\indent \indent \indent \indent $w(a, b) = \alpha(b) + 3 + \text{max}\left\{\displaystyle\sum_{\theta \in \Theta^+}\theta,\sum_{\theta \in \Theta^-}-\theta\right\} $\\
\indent \indent \indent $\Gamma(a) = \displaystyle\operatorname*{arg min}_{b \in B(a)}w(a,b)$\\
\indent \indent \indent \textbf{select} $\Delta(a) \in \displaystyle\operatorname*{arg min}_{\gamma \in \Gamma(a)}\left|\left\{ a \in A | \gamma \subset a \right\}\right|$\\
\indent \indent $D = \displaystyle \operatorname*{arg max}_{a \in A} w(a, \Delta(a))$\\
\indent \indent \textbf{select} $d\in D$\\
\indent \indent $K(\Delta(d)) = K\left(\Delta(d)\right) \cup \left(d \setminus  \Delta(d)\right)$\\
\indent \indent $A = A \setminus d$\\
\textsc{Output}: $K:\{\{i,j\})|1\leq i < j \leq N\}\rightarrow 2^{\{i|1\leq i\leq N\}}$
}}
\caption{Greedy algorithm for choosing which ancilla bits to use with each cubic term in reducing a cubic pseudo-Boolean to a quadratic one. The algorithm attempts to minimize the control precision of the reduced function. Given the function $\alpha$ that yields the coefficient of a term from the indices of its variables, the algorithm returns the function $K$ that yields the the set of indices of variables that together with the variables whose indices are passed to it form a cubic term to be reduced using the latter. See text for explanation.}
\end{figure}

The algorithm is initialized by setting $K(\{i,j\})$ to the empty set for every pair of variable indices $\{i,j\}$, and by collecting the triplet of variable indices $\{i,j,k\}$ for every 3-local term $\alpha_{ijk}x_ix_jx_k$ with a non-zero coefficient $\alpha_{ijk}$ into the set $A$. We also introduce the notation $B(a)$ for the set of three pairs of indices contained by a triplet of indices $a$, e.g. $B(\{i,j,k\})=\{\{i,j\},\{i,k\},\{j,k\}\}$. The remainder of the algorithm consists of a procedure for choosing a 3-local term (as represented by the set of indices of its variables $d$) and a pair of variables contained therein (also represented by their indices $\Delta(d)$) with which to collapse it, which is repeated until such a choice has been made for every term that we wish to collapse. Throughout, the set $A$ contains those terms for which the decision has not been made.

The repeated procedure is as follows: first, for every 3-local term $a \in A$ for a which a pair has not been chosen with which to collapse it and for every pair therein $b \in B(a)$, the cost of collapsing the term $a$ using that pair $b$ is calculated. 
The cost is defined as $w(a, b) = \alpha(b) + 3 + \text{max}\left\{\displaystyle\sum_{\theta \in \Theta^+}\theta,\sum_{\theta \in \Theta^-}-\theta\right\} $, where $\Theta$ is the set consisting of coefficients of terms that the pair $b$ has already been chosen to collapse and the coefficient of the current term $a$, and $\Theta^+$ and $\Theta^-$ are respectively the positive and negative elements thereof. Second, we choose a term $d$ and reduction pair $\Delta(d)$ that minimizes the cost$w$. For each term $a\in A$ we find the set of pair(s) $\Gamma(a)$ with the least cost of collapsing the term $a$. Note that here we follow the convention that argmin (argmax) returns the set of arguments for which the function has its minimum (maximum) value over the specified domain, i.e. $\displaystyle \operatorname*{arg min}_{x \in X} f(x) = \{ x \in X | f(x) = \operatorname*{min}_{x\in X} f(x)\}$.
\begin{figure}[h!]
\includegraphics[width=\columnwidth]{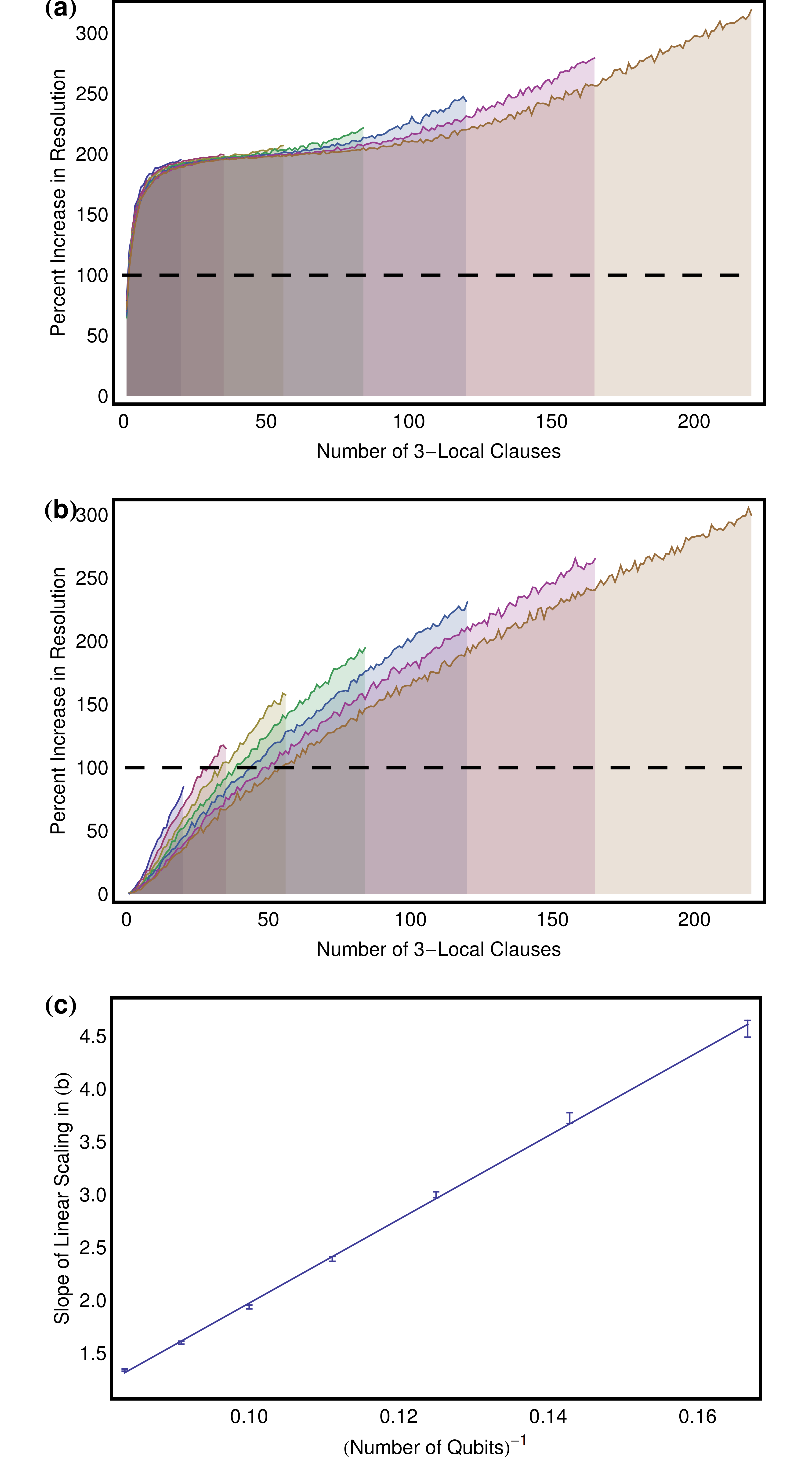}
\caption{\label{cp} Numerics were collected on randomly generated 3-local polynomial unconstrained binary optimization (\textsc{PUBO}) with $n$ logical qubits, $\lambda$ 3-local clauses, and $n$ choose 2, 2-local clauses. For each value of $n$, data were collected from 1,000 instances at every possible value of $\lambda$, i.e. $\left\{ \lambda \in \mathbb{Z} \mid 1 \leq \lambda \leq C^n_3\right\}$. Different colors indicate different values of $n$ ranging between 6 and 11 qubits.  Integer coefficients for each term were drawn at random from a flat distribution between $-8$ and 8. With these initial coefficients, a 100\% increase in control precision exhausts current D-Wave capabilities (this threshold is indicated with a dotted line). 3a: unoptimized application of reduction gadgets. 3b: application of our greedy algorithm indicating that increase in resolution is linear in $\lambda$. 3c: dependence of greedy algorithm's linear scaling in $\lambda$, suggesting that control precision is \emph{roughly} proportional to $\frac{\lambda}{n}$.}
\vspace{-5mm}
\end{figure}

If there is more than one such pair, we find which of those is contained in the fewest number of terms in $A$, those for which a choice has not yet been made. If there is then more than one such pair, a pair $\Delta(a)$ is chosen arbitrarily. Having found the minimum cost $w(a,\Delta(a))$ of each term $a\in A$, we find the set of terms with the minimum cost $D$ and choose one $d$ arbitrarily. Finally, we append the index in $d$ that is not in the reduction pair $\Delta(d)$ to $K(\Delta(d))$ and then remove the term $d$ from the set $A$ of terms for which a decision needs to be made. This procedure is repeated until a reduction pair has been chosen for every term, i.e. until $A$ is empty.

While we do not claim that this greedy algorithm is optimal, we present numerical evidence to show that it outperforms the default approach of selecting $K_{ij}$ in a non-systematic fashion. Figure 3 indicates that our technique significantly reduces the control precision cost of 3-local to 2-local reductions. For instance, with 11 qubits and 50 3-local terms, our approach requires approximately half the control precision that one would need with the default reduction strategy. Randomly choosing qubit pairs to collapse 3-local terms is the approach that many researchers (including the authors) have used in the past to encode problems into the D-Wave One device, even though the device is primarily control precision limited \cite{Perdomo-Ortiz2012}. Our results show that the expected increase in control precision is approximately proportional to $\lambda/n$, also known as the clause to variable ratio, $r$.

\section{Conclusion}

In this study, we have expanded the definition of an exact classical gadget and formalized the difficult problem of efficiently applying these tools. We introduced a novel and useful form of classical gadgets that uses multiple ancilla qubits to decrease the required control precision of compiling arbitrary problems. Using this new gadget we derived Eq.~\eqref{eq:controlPrecision}, a general expression for the optimal control precision of a 3-local to 2-local reduction. While exactly solving this equation appears extremely difficult, we introduced a simple greedy algorithm which significantly outperforms the status quo. For the problem of minimizing ancilla qubit requirements during 3-local to 2-local reduction, we demonstrated how to map the problem to \textsc{set cover} which allowed us to find minimal ancilla encodings with the use of \textsc{Integer Linear Programming}. We believe that these techniques will be very useful to anyone wishing to compile classical problems into realizable Hamiltonians for adiabatic quantum computation. We are working towards applying these new techniques for protein folding and related optimization problems of interest to chemistry and biophysics.

\section{Acknowledgements}
The authors thank David Tempel for helpful discussions. This research was sponsored by the United States Department of Defense. The views and conclusions contained in this document are those of the authors and should not be interpreted as representing the official policies, either expressly or implied, of the U.S. Government.

\section{Appendix}
\subsection{Reducing a quartic pseudo-Boolean to quadratic by mapping to \textsc{WMAXSAT}}
Here we show how the problem of reducing a quartic pseudo-Boolean to a quadratic one using the minimum number of ancilla bits can be recast as \textsc{Weighted Max-SAT} (\textsc{WMAXSAT}). An instance of \textsc{WMAXSAT} consists of a set of clauses, each of which is a disjunction of literals, and a function $w$ that assigns a non-negative weight to each clause; the problem is to find an assignment that maximizes the sum of the weights of clauses satisfied thereby. 

Consider an arbitrary 4-local term $x_i x_j x_k x_l$. It can be reduced to 2-local in two ways, both of which require two ancilla bits. The first way is to use two ancilla bits that each correspond to the conjunction of two computational bits. For example, the term can be reduced using the ancilla bits $x_{ij}$ and $x_{kl}$, which entails replacing the term $x_i x_j x_k x_l$ with $x_{ij}x_{kl}$ and adding the penalty functions $s(x_i,x_j,x_{ij})$ and $s(x_k,x_l,x_{kl})$, scaled by the appropriate factor. Similarly, the term can also be reduced using $x_{ik}$ and $x_{jl}$, or $x_{il}$ and $x_{jk}$. The second way is to use an ancilla bit corresponding to the conjunction of three bits, which requires a second ancilla bit.\footnote{No quadratic pseudo-Boolean $f(x,y,z,a)$ exists such that $f(x,y,z,a) = 0$ if $a=xyz$ and $f(x,y,z,a)\geq 1$  otherwise, which can be shown in a similar manner to that of the proof that the minimum coefficient in the penalty function for the conjunction of two variables is three.} For example, the term $x_i x_j x_k x_l$ can reduced to 2-local using the ancilla bits $x_{ij}^k$ and $x_{ij}$, where $x_{ij}^k$ corresponds to the conjunction of $x_{ij}$ and $x_k$.\footnote{Accordingly, just as the indices of the ancilla bit $x_{ij}$ were unordered, i.e. $x_{ij}=x_{ji}$, so are the subscript indices of the ancilla bit $x_{ij}^k$, i.e. $x_{ij}^k=x_{ji}^k$, though the distinction between subscript and superscript indices must be made. Though in reducing a single term the choice of which pair of computational bits to use for the intermediary ancilla bit is unimportant, when reducing several the same ancilla bit may be used as an intermediary for several ancilla bits each corresponding to the conjunction of three computational bits.} This entails replacing the term by $x_{ij}^k x_l$ and adding the penalty functions $s(x_i, x_j, x_{ij})$ and $s(x_{ij}, x_k, x_{ij}^k)$, scaled by the appropriate factor. There are twelve distinct ancilla bit pairs that can be used to reduce the term using the second way.

Now consider a quartic pseudo-Boolean 
\begin{align}
f(\mathbf{x}) &	= \alpha_0 + \sum_{1\leq i \leq N} \alpha_i x_i + \sum_{1\leq i<j\leq N} \alpha_{ij} x_i x_j \\
& \quad {} + \sum_{1\leq i < j < k \leq N} \alpha_{ijk} x_i x_j x_k + \sum_{1\leq i < j < k < l \leq N} \alpha_{ijkl} x_i x_j x_k x_l \nonumber
\end{align}
that we would like to reduce to quadratic. Let $T_3$ and $T_4$ be sets of the sets of indices of the variables in the 3-local and 4-local terms with non-zero coefficients, respectively, i.e.
\begin{align} 
T_3 &= \{ \{ i,j,k \} \subset \{1, \ldots, N\} | \alpha_{ijk} \neq 0 \} \\
\intertext{and}
T_4 &= \{ \{ i,j,k,l \} \subset \{1, \ldots, N\} | \alpha_{ijkl} \neq 0 \}.
\end{align}
For each ancilla bit $x_{ij}$ that represents a conjunction of two computational bits, we introduce a Boolean variable $r_{ij}\in \{\textsc{true}, \textsc{false}\}$ that represents its actual use. For each triplet of computational bits $\{x_i,x_j,x_k\}$, we introduce a Boolean variable $r_{ijk}\in \{\textsc{true}, \textsc{false}\}$ corresponding to the use of an ancilla corresponding to their conjunction, regardless of which intermediate ancilla bit was used. While the choice of intermediate ancilla bit must be made when doing the reduction, the minimum set of ancilla bits used in a reduction cannot contain two distinct ancilla bits corresponding to the conjunction of the same three ancilla variables and so here there is no need to make the distinction. Let
\begin{align}
R_2 &= \left\{r_{ij} | \{i,j\}\subset \bigcup_{t\in T_3 \cup T_4}t\right\},\\
R_3 &= \left\{r_{ijk} | \{i,j,k\} \subset \bigcup_{t\in T_4} t\right\},\\
\intertext{and}
R &= R_2 \cup R_3.
\end{align}

There are three sets of clauses that must be included. First, the goal is to minimize the number of ancilla bits used in the reduction, and so for each variable representing the use of a unique ancilla bit we include the single-literal clause consisting of its negation, and assign to each such clause a weight of 1:
\begin{equation}
\mathcal F_1 = \{ (\overline{r_{ij}} | \{i,j\} \in R_2 \} \cup
\{ (\overline{r_{ijk}} | \{i,j,k\}\in R_3 \} 
\end{equation}
and $w(C) = 1$ for every $C \in \mathcal F_1$. This first set consists of so-called soft clauses. The remaining two sets of clauses $\mathcal F_2$ and $\mathcal F_3$ consist of hard clauses, those that must be satisfied. This is ensured by assigning to every hard clause a weight greater than the sum of the weights of al the soft clauses. Here, we set $w(C)=|\mathcal F_1| + 1 = |R| + 1 $ for every $C \in \mathcal F_2 \cup \mathcal F_3$. Note that $|R|\leq \binom{N}{3} + \binom{N}{2} = \frac{n(n^2-1)}{6}$.

Second, we must ensure that for each ancilla bit used  that corresponds to the conjunction of three computational bits there is at least one intermediate ancilla bit that can be used in its construction, i.e.
\begin{equation}
(r_{ijk} \to (r_{ij} \lor r_{ik} \lor r_{jk})) \equiv 
(\overline{r_{ijk}} \lor r_{ij} \lor r_{ik} \lor r_{jk}).
\end{equation}
Let
\begin{equation}
\mathcal F_2 =
\{ (\overline{r_{ijk}} \lor r_{ij} \lor r_{ik} \lor r_{jk}) 
   | \{i,j,k\} \in R_3\}.
\end{equation}
Third, we must ensure that the set of ancilla bits used reduces all the cubic and quartic terms. A cubic term $x_ix_jx_k$ can be reduced using $x_{ij}$, $x_{ik}$, or $x_{jk}$, i.e. if $(r_{ij} \lor r_{ij} \lor r_{jk})$. Note that while an ancilla bit corresponding to the term itself can be used to reduce it to 1-local, that ancilla bit can only be constructed using one of the three ancilla bits mentioned, and any one of those three is sufficient to reduce the term to quadratic. A quartic term $x_i x_j x_k x_l$ can be reduced using one of twelve ancilla bits (though each requires an intermediary). These twelve can be partitioned into four triplets by the triplet of variables whose conjunction they correspond to, i.e. by the Boolean variable that represents the use of any one. Thus the quartic term can be reduced to quadratic if $(r_{ijk} \lor r_{ijl} \lor r_{ikl} \lor r_{jkl})$. It can also be reduced using two ancilla bits that correspond to the conjunctions of disjoint pairs of computational bits, i.e. if 
$((r_{ij} \land r_{kl}) \lor (r_{ik} \land r_{jl}) \lor (r_{il} \land r_{jk}))$. These clauses must be written in conjunctive normal form:
\begin{align}
((r_{ij} \land r_{kl}) \lor (r_{ik} \land r_{jl}) \lor (r_{il} \land r_{jk}) 
\lor r_{ijk} \lor r_{ijl} \lor r_{ikl} \lor r_{jkl}) \nonumber\\
\equiv  
\bigwedge_{\substack{
y_1 \in \{ r_{ij}, r_{kl} \} \\ 
y_2 \in \{ r_{ik}, r_{jl} \} \\ 
y_3 \in \{ r_{il}, r_{jk} \} }}
(y_1 \lor y_2 \lor y_3 \lor r_{ijk} \lor r_{ijl} \lor r_{ikl} \lor r_{jkl}).\nonumber
\end{align}
Let
\begin{align}
\mathcal F_3 &= \{ (r_{ij} \lor r_{ij} \lor r_{jk}) | \{i,j,k\} \in T_3 \}
\\
& \quad {} \cup \bigcup_{\{i,j,k,l\} \in T_4} \bigcup_{\substack{
y_1 \in \{ r_{ij}, r_{kl} \} \\ 
y_2 \in \{ r_{ik}, r_{jl} \} \\ 
y_3 \in \{ r_{il}, r_{jk} \} }}
(y_1 \lor y_2 \lor y_3 \lor r_{ijk} \lor r_{ijl} \lor r_{ikl} \lor r_{jkl}). \nonumber
\end{align}
Finally, let $\mathcal F = \mathcal F_1 + \mathcal F_2 + \mathcal F_3$. The \textsc{WMAXSAT} instance is specified by $\mathcal F$ and 
$w(C)= \begin{cases} 
1 & C \in \mathcal F_1 \\
|R|+1 & C \in \mathcal F_2 \cup F_3
\end{cases}$.

\subsection{Maximum number of ancilla bits needed to reduce a cubic pseudo-Boolean to quadratic}

We prove here that the minimum number of ancilla variables needed to reduce all 3-local terms over $n$ variables to 2-local is 
$\left\lfloor \frac{(n-1)^2}{4} \right\rfloor$, and therefore that the minimum number of ancilla variables needed to reduce any set of 3-local terms over $n$ variables is upper-bounded by the same.

The basis of the proof is Mantel's Theorem: A triangle-free graph with $n$ vertices can have at most $\left\lfloor \frac{n^2}{4}\right\rfloor$ vertices.\cite{Bollobas1998} We identify a set of ancilla bits $A$ used to reduce locality with the edge set $E(A)$ of a graph $G(A)$ whose vertices $V=\{v_i|1\leq i\leq N\}$ correspond to the computational variables and in which there is an edge between any two vertices $v_i$ and $v_j$ if and only if the ancilla bit $x_{ij}$ representing the conjunction of the corresponding computational bits $x_i$ and $x_j$ is used.\footnote{In reducing a cubic pseudo-Boolean to a quadratic, only ancilla bits of this type are needed.} The set of ancilla bits $A$ can be used to reduce all possible 3-local terms if and only if for every set of three computational bits there is at least one ancilla bit in $A$ corresponding to the conjunction of any two. In graph-theoretic terms, $A$ can be used to reduce all 3-local terms if and only if every possible triangle in the complete graph with the same the vertex set $V$ contains at least one edge in $E(A)$, or equivalently if the complement $E^C(A)$ of $E(A)$ is triangle-free. Suppose that the set of ancilla bits $A$ reduces all 3-local terms. Then by Mantel's Theorem $|E^C(A)|\leq \lfloor \frac{N^2}{4}\rfloor$. Because $|E(A)|+|E^C(A)|=\binom N2$, this yields
\begin{align}
|E(A)| &= \binom N2 - |E^C(A)| \\
& \geq \binom N2 - \left\lfloor \frac{N^2}{4} \right\rfloor.
\end{align} 
Let $N=2m+b$, where $m=\left\lfloor \frac N2 \right\rfloor \in \mathbb Z$ and $b = N - 2m \in \{0, 1\}$. Then
\begin{align}
|E(A)| &\geq \binom{2m+b}{2} - \left\lfloor \frac{(2m+b)^2}{4} \right\rfloor\\
& = \frac{(2m+b)(2m+b-1)}{2} - \left\lfloor m^2 + mb + \frac{b^2}{4} \right \rfloor\\
&= 2m^2+2mb-m+\frac{b^2-b}{2} - (m^2 + mb)\\
&= m^2 + mb - m \\
&= \left \lfloor m^2 + mb -m +\frac{b^2-2b+1}{4}  \right\rfloor \\
&= \left \lfloor \frac{(2m+b-1)^2}{4} \right\rfloor \\
&= \left \lfloor \frac{(N-1)^2}{4} \right\rfloor.
\end{align}
Furthermore, by construction we show that the minimal set reaches this bound. Let $E = \{ \{ v_i, v_j\} | (1 \leq i < j \leq \lceil N/2\rceil) \lor (\lceil N/2\rceil + 1\leq i<j\leq N\}$. That is, partition the vertices into sets of as equal size as possible and include an edge between every pair within each set. Let $N=2m+b$ as above. The total number of edges constructed in this way is 
\begin{align}
\binom{\lceil N/2 \rceil}{2}+\binom{\lfloor N/2 \rfloor}{2}  &= \binom{m+b}{2} + \binom m2 \\
&= \frac{(m+b)(m+b-1)}{2} + \frac{m(m-1)}{2}\\
&= m^2 + mb - m\\
&= \lfloor \frac{(N-1)^2}{4}\rfloor.
\end{align}

\bibliographystyle{ieeetr}
\bibliography{./library}

\end{document}